\documentclass[runningheads,a4paper]{llncs}

\usepackage{amssymb}
\usepackage[english]{babel}
\usepackage{graphicx}
\usepackage{algorithmic}
\usepackage{algorithm}
\usepackage{color}

\usepackage{bbding}

\usepackage{url}
\newcommand{\keywords}[1]{\par\addvspace\baselineskip
\noindent\keywordname\enspace\ignorespaces#1}

\begin{document}
\pagestyle{empty}

\mainmatter  

\title{How to define \emph{co-occurrence} in different domains of study?}


%
%
\author{Mathieu Roche $^{1,2}$}%


\authorrunning{Roche}
\institute{Cirad, TETIS, F-34398 Montpellier, France \\
\and
TETIS, Univ. Montpellier, AgroParisTech, Cirad, CNRS, Irstea, Montpellier, France \\
\url{mathieu.roche@cirad.fr}\\
\url{http://textmining.biz/Staff/Roche}
}

\maketitle

\thispagestyle{empty}

\begin{abstract}

This position paper presents a comparative study of  \emph{co-occurrences}. Some similarities and differences in the definition exist depending on the research domain (e.g. linguistics, NLP, computer science). This paper discusses these points, and deals with the methodological aspects in order to identify  \emph{co-occurrences} in a multidisciplinary paradigm.
\end{abstract}

\section{Introduction}

Determining \emph{co-occurrences} in corpora is challenging for different applications such as classification, translation, terminology building, etc. More generally, \emph{co-occurrences}  can be identified with all types of data, e.g. databases \cite{CAO2007343}, texts \cite{DBLP:conf/iis/RocheAMK04}, images \cite{Verma:2015:LEC:2798736.2798879}, music \cite{Ghosal2011}, video \cite{Jeon2005}, etc. \\

The \emph{co-occurrence} concept has different definitions depending on the research domain (i.e. linguistics, NLP, computer science, biology, etc.). This position paper reviews the main definitions in the literature and discusses similarities and differences according to the domains. This type of study can be crucial in the context of data science, which is geared towards developing a multidisciplinary paradigm for data processing and analysis, especially textual data. \\

Here the \emph{co-occurrence} concept related to textual data is discussed. 
Note that before their validation by an expert, co-occurrences of words are often considered as \emph{candidate terms}. \\

First, Section \ref{definition} of this paper details the different definitions of \emph{co-occurrence} according to the studied domains. Section \ref{discussion} discusses and compares these different aspects based on their intrinsic definition but also on the associated methodologies in order to identify them. Finally, Section \ref{conclusion} lists some perspectives.

\section{\emph{Co-occurrence} in a multidisciplinary context} \label{definition}

\subsection{Linguistic viewpoint}

In linguistics, one notion that is widely used to define the term is called \emph{lexical unit} \cite{Lederer69} and  \emph{polylexical expression} \cite{Gross96}. 
The latter represents a set of words having an autonomous existence, which is also called \emph{multi-word expression} \cite{Sag:2002:MEP:647344.724004}. \\

In addition, several linguistics studies use the \emph{collocation} notion. 
\cite{Clas94} gives two properties defining a collocation. First, collocation is defined as a group of words having an overall meaning that is deducible from the units (words). For example, \emph{climate change} is considered as a collocation because the overall meaning of this group of words can be deduced from both words \emph{climate} and \emph{change}. On the other hand, the expression \emph{to rain cats and dogs} is not a collocation because its meaning cannot be deduced from each of the words; this is called a \emph{fixed expression} or an \emph{idiom}. \\

A second property is added by \cite{Clas94} to define a collocation. The meaning of the words that make up the collocation must be limited. For example, \emph{buy a dog} is not a collocation because the meaning of \emph{buy} is not limited.

\subsection{NLP viewpoint}

In the natural language processing (NLP) domain, the \emph{co-occurrence} notion refers to the general phenomenon where words are present together in the same context. More precisely, several principles are used that take contextual criteria into account. \\

First, the terms or phrases \cite{Bourigault:1992:SGA:992383.992415,Daille:1994:TAE:991886.991975} can respect syntactic patterns (e.g. adjective noun, noun noun, noun preposition noun, etc.). Some examples of extracted phrases (i.e. \emph{syntactic co-occurrences}) are given in Table \ref{example}. \\

In addition, methods without linguistic filtering are also conventionally used in the NLP domain by extracting $n$-grams of words (i.e. \emph{lexical co-occurrences})  \cite{DBLP:journals/nle/MassungZ16,e46325429cf245b58edc6f687b3aac1e}. $n$-grams are contiguous sequences of $n$ words extracted from a given sequence of text  (e.g. the bi-grams\footnote{$n$-grams with $n=2$.} $x$ $y$ and $y$ $z$ are associated with the text $x$ $y$ $z$).
$n$-grams that allow gaps are called skip-$n$-grams (e.g. the skip-bi-grams $x$ $y$, $x$ $z$, $y$ $z$ are related to the text $x$ $y$ $z$). 
Skip-gram model is an efficient method for learning high-quality distributed vector representations that capture a large number of precise syntactic and semantic word relationships \cite{Mikolov:2013:DRW:2999792.2999959}. 
Some examples of $n$-grams and skip-$n$-grams are given in Table  \ref{example}.\\

After summarizing the term notion in the NLP domain, the following section discusses these aspects in the computer science context, particularly in data mining. Note that the NLP domain may be considered as being located at the linguistics and computer science interface.

\begin{table}[h]
\begin{center}
   \begin{tabular}{|c|c|}
   \hline
     \multicolumn{2}{|c|}{}    \\ 
            \multicolumn{2}{|c|}{}    \\ 
       \multicolumn{2}{|c|}{{\bf Sentence} (input) } \\
        \multicolumn{2}{|c|}{}    \\ 
     \multicolumn{2}{|c|}{\it With climate change the water cycle is expected to undergo significant change.}    \\ 
      \multicolumn{2}{|c|}{}    \\ 
      \hline
        \multicolumn{2}{|c|}{}    \\ 
       \multicolumn{2}{|c|}{{\bf Candidates} (output) } \\
        \multicolumn{2}{|c|}{}    \\

   \hline
     {\bf Phrases} & {\it climate change } \\
        (noun noun, adjective-noun) & {\it water cycle, significant change } \\
     \hline
    {\bf bi-grams of words} &  {\it With climate, climate change, change the, the water, }\\
    &  {\it water cycle, cycle is, is expected, expected to, } \\
    & {\it  to undergo, undergo significant, significant change} \\
   \hline
   {\bf 2-skip-bi-grams} & {\it With climate, With change, With the, }\\
& {\it climate change, climate the, climate water, } \\
& {\it change the, change water, change cycle, } \\
& {\it the water, the cycle, the is, } \\
& {\it water cycle, water is, water expected, } \\
& {\it cycle is, cycle expected, cycle to, } \\
& {\it is expected, is to, is undergo, } \\
& {\it expected to, expected undergo, expected significant, } \\
& {\it to undergo, to significant, to change, } \\
& {\it undergo significant, undergo change, }\\
& {\it significant change} \\
   \hline

\end{tabular}
\end{center}
\caption{Examples of candidates extracted with different NLP techniques. \label{example}}
\end{table}

\subsection{Computer science viewpoint}

In the data mining domain, co-occurring items are called \emph{association rules}  \cite{Agrawal:1994:FAM:645920.672836,Yin2011} and they could be candidates for construction or enrichment of terminologies \cite{DBLP:conf/otm/Di-JorioBFLT08}. \\

In the data mining context, the list of items corresponds to the set of available articles. With textual data, items may represent the words present in sentences, paragraphs, or documents \cite{Amir2005,DBLP:conf/f-egc/RabatelLPSSRL08}. A transaction is a set of items. A set of transactions is a learning set used to determine association rules. \\

Some extensions of association rules are called \emph{sequential patterns}. They take into account a certain order of extracted elements \cite{Jaillet:2006:SPT:1165444.1165446,Serp08} with an enriched representation related to textual data as follows: 
\begin{itemize}
\item \emph{objects} represent texts or pieces of texts,
\item \emph{items} are the words of a text,
\item \emph{itemsets} represent sets of words present together within a sentence, paragraph or document,
\item \emph{dates} highlight the order of sentences within a text. \\
\end{itemize}

There  are  several  algorithms  for discovering association rules and sequential patterns.  One of the  most  popular   is  Apriori, which  is  used  to  extract  frequent  itemsets  from  large  databases.  The Apriori algorithm \cite{Agrawal:1994:FAM:645920.672836} finds frequent itemsets where $k$-itemsets  are  used  to  generate  $k+1$-itemsets.  \\

Association rules and sequential patterns of words are often used in text mining for different applications, e.g. terminology enrichment \cite{DBLP:conf/otm/Di-JorioBFLT08}, association of concept instances \cite{BERRAHOU2017115,DBLP:conf/f-egc/RabatelLPSSRL08}, classification \cite{Jaillet:2006:SPT:1165444.1165446,Serp08}, etc.


\section{Discussion: comparative study of definitions and approaches} \label{discussion}

This section proposes a comparison of : (i) \emph{co-occurrence} definitions (see Section \ref{candidate}), (ii) automatic methods in order to identify them (see Section \ref{methodextraction}).
This section highlights some similarities and differences between domains.

\subsection{\emph{Co-occurrence} extraction} \label{candidate}

The general definition of \emph{co-occurrence} is finally close to \emph{association rules} in data mining domain.  
Note that the integration of windows\footnote{Association Rule with Time-Windows (ARTW) \cite{Yin2011}.} in the association rule or sequential pattern extraction process enables us to have similarity with skip-$n$-gram extraction. \\

The integration of syntactic criteria makes it possible to extract more relevant candidate terms (see Table \ref{example}). Such information is typically taken into account in NLP to extract terms from general or specialized domains \cite{JiangDTCX12,Lossio-Ventura:2016:BTE:2890328.2890347,Nenadic:2003:TMB:952532.952553,Roche2017valocarn}. \\

Table \ref{example} highlights relevant terms extracted using linguistic patterns (e.g. {\it climate change, water cycle, significant change}). The use of linguistic patterns tends to improve precision values. Generally other methods such as skip-bi-grams return lower precision, i.e. many extracted candidates are irrelevant (e.g. {\it climate the}). But this kind of method enables extraction of some relevant terms not found with linguistic patterns (e.g. {\it cycle expected}); then the recall can be improved. \\

Table \ref{synthese} presents research domains related to different types of candidates, i.e. collocations, polylexical expressions, phrases, $n$-grams, association rules, sequential patterns. \\

Table \ref{synthese2} summarizes the main criteria described in the literature. Note that the extraction is more flexible and automatic when there are fewer criteria. In this table, two types of information are associated with the different criteria. The first one (marked with $\checkmark$) designates the characteristics given by the \emph{co-occurrence} definitions. The second type of information (marked with $\bigstar$) represents characteristics that are implemented in many extensions of the state-of-the-art. \\

\begin{table}[h]
\begin{center}
   \begin{tabular}{|c|c|}
   \hline
    Definitions & Domains    \\
   \hline
  {\bf Collocations} & L   \\
  \hline
 {\bf Polylexical expressions} & L + NLP  \\
   \hline
    {\bf Phrases} & NLP \\
     \hline
    {\bf n-grams} & NLP + CS \\
   \hline
   {\bf Association rules} & CS \\
   \hline
   {\bf Sequential patterns} & CS \\
   \hline
\end{tabular}
\end{center}
\caption{Summary of the main domains associated with expressions (L: linguistics, NLP: natural language processing, CS: computer science). \label{synthese}}
\end{table}

\begin{table}[h]
     \begin{center}
   \begin{tabular}{|c|c|c|c|c|}
   \hline
     & Ordered  & Sequences  &  Morpho-syntactic  &   Semantic  \\
     & sequences & with gaps & information &   information \\
   \hline
  {\bf Collocations} & $\checkmark$ & &  $\checkmark$ & $\bigstar$ \\
  \hline
 {\bf Polylexical expressions} &  $\checkmark$  & &  $\checkmark$  & \\
   \hline
    {\bf Phrases}  & $\checkmark$  & &  $\checkmark$  & \\
     \hline
    {\bf $n$-grams}  & $\checkmark$  & $\bigstar$ &  & \\
   \hline
   {\bf Association rules}  &  & $\checkmark$ &  & \\
   \hline
   {\bf Sequential patterns}  & $\checkmark$  & $\checkmark$  &  & \\
   \hline
\end{tabular}
\end{center}
\caption{Summary of the main criteria associated with \emph{co-occurrence} identification. $\checkmark$  represents the respect of the criterion by definition. $\bigstar$ is present when extensions are currently used in the state-of-the-art. \label{synthese2}}
\end{table}


Table \ref{synthese2} shows that the semantic criterion is seldom associated with \emph{co-occurrence} definitions. This criterion is however taken into account in linguistics. For example, semantic aspects are taken into account in several studies \cite{Heid98,Laurens99,Melcuk-etal84-99}. 
In this context \cite{Melcuk-etal84-99} introduced \emph{lexical functions} rely on semantic criteria to define the relationships between collocation units.
For instance, a given relation can be expressed in various ways between the arguments and their values, like \emph{Centr (the center, culmination of)} that returns different meanings\footnote{http://people.brandeis.edu/$\sim$smalamud/ling130/lex\_functions.pdf}:
 \begin{itemize}
\item {\it Centr(crisis) = the peak }
\item {\it Centr(desert) = the heart }
\item {\it Centr(forest) = the thick }
\item {\it Centr(glory) =  summit }
\item {\it Centr(life) = prime  }
 \end{itemize}

In the data mining domain, semantic information is used in two main directions. The first one involves filtering the results if they respect certain semantic information (e.g. phrases or patterns where a word is an instance of a semantic resource). Other methods involve semantic resources in the knowledge discovery process, i.e. the extraction is driven by semantic information \cite{BERRAHOU2017115}.  \\

In recent studies in the NLP domain, the semantic aspects are based on word embedding, which provides a dense representation of words and their relative meanings \cite{Ganguly:2015:WEB:2766462.2767780,Zamani:2017:RWE:3077136.3080831}. \\

Finally, note that several types of \emph{co-occurrence} are often used in different domains. For example, polylexical expressions are commonly used in NLP and also in linguistics. In addition, $n$-grams is currently used in NLP and computer science domains. For example, $n$-grams of words are often used to build terminologies (NLP domain) but also as features for machine learning algorithms (computer science domain) \cite{e46325429cf245b58edc6f687b3aac1e}. \\

Table  \ref{synthese3} summarizes the main types of criteria (i.e. statistic, morpho-syntactic, and semantic) used for extracting \emph{co-occurrences} according to the research domains considered in this paper.

\begin{table}[h]
     \begin{center}
   \begin{tabular}{|c|c|c|c|c|}
   \hline
     &  Statistic  &  Morpho-syntactic  &   Semantic  \\
     &  information & information &   information \\
   \hline
  {\bf Linguistics } &  & $\checkmark$   &   $\bigstar$    \\
  \hline
 {\bf NLP} &  $\checkmark$  & $\checkmark$ &  $\bigstar$   \\
   \hline
    {\bf Data mining}  & $\checkmark$  & $\bigstar$ &  $\bigstar$   \\
   \hline
\end{tabular}
\end{center}
\caption{Summary of the main criteria associated with research domains. $\checkmark$ represents the respect of the criterion for extracting \emph{co-occurrences} from textual data. $\bigstar$ is present when extensions are currently used in the state-of-the-art.  \label{synthese3}}
   
\end{table}

After presenting the characteristics associated with the \emph{co-occurrence} notion in a multidisciplinary context, the following section compares the methodological viewpoints to identify these elements according to the domains.


\subsection{Ranking of \emph{co-occurrences}}  \label{methodextraction}

\emph{Co-occurrence} identification by automatic systems is generally based on the use of quality measures and/or algorithms. This section provides two illustrative examples that show similarities between approaches according the domains.  \\

\subsubsection{ \emph{Mutual Information} and  \emph{Lift measure}}  
~ \\

First the use of specific statistical measures from different domains is highlighted. This paragraph focuses on the study of Mutual Information (MI). 
This measure is often used in the NLP domain to measure the association between words \cite{Church:1990:WAN:89086.89095}. 
MI (see formula (\ref{IM})) compares  the  probability  of observing $x$ and  $y$ together (joint probability) with the probability of observing $x$ and $y$ independently (chance) \cite{Church:1990:WAN:89086.89095}.

\begin{equation}
I(x) = log_2 \frac{P(x,y)}{P(x)P(y)} 
\label{IM}
\end{equation}

~ \\

In general, word  probabilities $P(x)$ and $P(y)$ correspond to the number of observations of $x$ and $y$ in a corpus, normalized by the size of the corpus.  Some extensions of $MI$ are also proposed. The algorithm PMI-IR (Pointwise Mutual Information and Information Retrieval) described in \cite{turney01mining} queries the Web via the AltaVista search engine to determine appropriate synonyms for a given query.  For a given word, denoted $x$, PMI-IR chooses a synonym among a given list. These selected terms, denoted $y_i$, $i \in [1,n]$, correspond to TOEFL questions. 
The aim is to compute the $y_i$ synonym that gives the best score. 
To obtain scores, PMI-IR uses several measures based on the proportion of documents where both terms are present. Turney's formula is given below  (\ref{formule_turney1}): It is one of the basic measures used in \cite{turney01mining}. 
It is inspired from MI described in \cite{Church:1990:WAN:89086.89095}. With this formula (\ref{formule_turney1}), the proportion of documents containing both $x$ and $y_i$ (within a $10$ word window) is calculated, and compared with the number of documents containing the word $y_i$. The higher this proportion, the more $x$ and $y_i$ are seen as synonyms.

\begin{eqnarray} \label{formule_turney1}
score(y_i) = \frac{nb(x\mbox{
}NEAR\mbox{ }y_i)}{nb(y_i)}
\end{eqnarray}
\begin{itemize}
\item $nb(x)$ computes the number of documents containing the word $x$ (i.e. $nb$ corresponds to number of webpages returned by search engines),
\item $NEAR$  (used in the 'advanced research' field of AltaVista) is an operator that identifies if two words are present in a $10$ word wide window. \\
\end{itemize}

This kind of web mining approach is also used in many NLP applications, e.g. (i) computing the relationship between \emph{host} and \emph{clinical sign}  for an epidemiology surveillance system \cite{DBLP:journals/ijaeis/ArsevskaRHCFLD16}, (ii) computing the dependency of words of acronym definitions for word-sense disambiguation tasks \cite{DBLP:journals/informaticaSI/RocheP10}. \\

The probabilities are generally symmetric (i.e. $P(x,y) = P(y, x)$), while the original MI measure is also symmetric. 
But the association ratio applied in the NLP domain is not symmetric, i.e. the occurrence number of pairs of words "$x$ $y$" and  "$y$ $x$" generally differ. Moreover the meaning and relevance of phrases should differ according to the word order in a text, e.g. \emph{first lady} and \emph{lady first}. \\

Finally, MI is very close to the \emph{lift} measure \cite{Brin:1997:BMB:253260.253327,Ventura2016,Azevedo:2007:CRM:1421665.1421715} in data mining. 
This measure identifies relevant association rules (see formula (\ref{lift})). 
The lift measure evaluates the relevance of co-occurrences only (not implication) and how $x$ and $y$ are independent \cite{Azevedo:2007:CRM:1421665.1421715}.

\begin{equation}
lift(x \rightarrow y) = \frac{conf(x \rightarrow y)}{sup(y)} 
\label{lift}
\end{equation}

This measure is based on both \emph{confidence} and \emph{support} criteria, which in turn are based on association rule ($x \rightarrow y$) identification. Support is an indication of how frequently the itemset appears in the dataset. 
Confidence is a standard measure that estimates the probability of observing $y$ given $x$ (see formula \ref{conf}).

\begin{equation}
conf(x \rightarrow y) = \frac{sup(x \cup y)}{sup(x)} 
\label{conf}
\end{equation}

Note that other quality measures of the data mining domain, such as \emph{Least contradiction} or \emph{Conviction} \cite{Lallich2007}, could be tailored to deal with textual data.

\subsubsection{ \emph{C-value} and  \emph{closed itemset}}  
~ \\

Another example is the methodological similarities associated with different approaches. For example, the C-value approach \cite{Frantzi2000} used in the NLP domain \cite{Lossio-Ventura:2016:BTE:2890328.2890347,JiangDTCX12} favors terms that do not appear to a significant extent in longer terms. For example, in a specialized corpus related to ophthalmology, \cite{Frantzi2000}  show that a more general term such as \emph{soft contact} is irrelevant, whereas a longer and therefore more specific term such as \emph{soft contact lens} is relevant.
This kind of measure is particularly relevant in the biology domain \cite{Lossio-Ventura:2016:BTE:2890328.2890347,JiangDTCX12}. \\ 

In addition, in the computer science domain (i.e. data mining), the notion of \emph{closed itemset} is finally very close to the C-value approach. In this context, a frequent itemset is considered as closed if none of its supersets\footnote{A superset is defined with respect to another itemset, for example \{M1, M2, M3\} is a superset of \{M1, M2\}. B is superset of A if card(A) $<$ card(B) and A $\subset$ B.} has the same support (i.e. frequency).  \\
  

This section and both illustrative examples confirm the importance of having a real multidisciplinary viewpoint on the methodological aspects, in order to build scientific bridges and thus contribute to the development of the emerging data science domain. \\



\section{Conclusion and Future Work}  \label{conclusion} 

This position paper proposes a discussion on similarities as well as differences in the definition of \emph{co-occurrence} according to research domains (i.e. linguistics, NLP, computer science). The aim of this position paper is to show the bridges that exist between different domains.  \\

In addition, this paper highlights some similarities in the methodologies used in order to identify \emph{co-occurrences} in different domains. We could extend the discussion to other domains. For example, methodological transfers are currently applied between bioinformatics and NLP. For example, the use of edition measures (e.g. Levenshtein distance) for sequence alignment tasks (bioinformatics) \emph{v.s.} string comparison (NLP). 

~ \\

\subsection*{Acknowledgments} 

This work is funded by the SONGES project (Occitanie and FEDER) -- Heterogeneous Data Science (\url{http://textmining.biz/Projects/Songes}).

~ \\

\bibliographystyle{splncs}
\bibliography{biblio}


\end{document}